\documentclass[aps,showpacs,preprint]{revtex4}%
\usepackage{amsfonts}
\usepackage{amsmath}
\usepackage{amssymb}
\usepackage{graphicx}
\usepackage{makeidx}
\usepackage{epsfig}
\usepackage{psfig}
\usepackage{subfigure}%
\begin{document}
\title{Generation of Multi-atom Dicke States with Quasi-unit Probability through the
Detection of Cavity Decay}
\author{Yun-Feng Xiao}
\author{Zheng-Fu Han}
\author{Jie Gao}
\author{Guang-Can Guo}
\affiliation{Key Laboratory of Quantum Information, University of Science and Technology of
China, Hefei 230026, People's Republic of China}

\pacs{03.67.Hk, 42.50.Pq}

\begin{abstract}
We propose a scheme to create universal Dicke states of n largely detuned
atoms through detecting the leaky photons from an optical cavity. The
generation of entangled states in our scheme has quasi-unit success
probability, so it has potential practicability based on current or near
coming laboratory cavity QED technology.

\end{abstract}
\maketitle

Entanglement plays an important role in the fields of quantum theory and
quantum information processing (QIP). It is not only used to test quantum
mechanics against local hidden variable theory \cite{bell}, but also holds the
keys of the applications of QIP including computation \cite{grover},
communication \cite{nielsen}, and cryptography \cite{ekert}. There are many
theoretical and practical schemes to entangle two particles or
multi-particles, such as spontaneous parametric down converters
\cite{bouwmeester}, linear ion trap \cite{sackett}, atomic ensembles
\cite{duan}, and cavity QED \cite{rauschenbeutel}. Among them, the schemes
based on cavity QED attract persistent interest in the experimental
realization because long-lived states in high-$Q$ cavities provide a promising
tool for creating entanglement and superposition, and also possible for
implementation of quantum computing \cite{xiao}. Entangled states for
two-level atoms in cavity QED have been observed for
\cite{rauschenbeutel,hagley} in experiment to date.

One of the main obstacles for the implementation of quantum information in
cavity QED, including preparation of multi-atom entanglement, is the
decoherence of the atoms and\ cavity fields. As potential solutions, largely
detuned atoms in the cavities \cite{zheng} and dissipation-assisted
conditional quantum evolution via the detection of cavity decay \cite{many}
have been proposed to date. On the one hand, when the atoms have large
detuning with the cavity modes, the atomic populations in the excited states
are very small and thus atomic spontaneous emission (described by the rate
$\gamma_{s}$) can be neglected. In other words, the atomic excited states can
be eliminated due to large detuning and atoms evolve only in their ground
state space. On the other hand, cavity decay is also considered as a useful
ingredient, not a destructive factor, since the idea was proposed by Plenio
and Knight \cite{knight}.

Recently, Hong and Lee proposed a scheme to generate two-atom entangled state
in a cavity quasideterministicly \cite{hong}. In their paper, two three-level
atoms should couple resonantly with two different polarized cavity modes.
Therefore, the atomic\ spontaneous emission is a significant reason to
decrease the success probability, especially when the trials have been done
for several times. Here, we propose an extended and improved scheme to
generate Dicke states \cite{mandel} of arbitrary n among N trapped largely
detuned atoms with quasi-unit probability. Furthermore, we extend our
conclusion to generate a non-trivial subset of Dicke states of the n atoms
step by step. The Dicke states are defined as $\left\vert n,m\right\rangle
_{Dicke}=C\left(  n,m\right)  \left(  s_{1}\right)  ^{m}\left(  s_{0}\right)
^{n-m}\otimes_{j=1}^{n}\left\vert e\right\rangle _{j}$ \cite{mandel}. Here,
the collective operators $s_{k}$ ($k=0,1$) are defined as $s_{k}=%
{\displaystyle\sum\nolimits_{j=1}^{n}}
\left\vert k\right\rangle _{jj}\left\langle e\right\vert $, and the
normalization coefficient $C\left(  n,m\right)  =1/\sqrt{n!m!\left(
n-m\right)  !}$. The muti-atom Dicke states and the GHZ states in general
belong to different classes of entangled states, and Dicke states are
relatively more immune to the influence of noise \cite{dur}. Dicke states have
many interesting applications in QIP and in high-precision measurements
\cite{cabello}.

In the current practical cavity QED system, optical cavity always has a finite
quality factor, and the coherent coupling rate $g\left(  r\right)  $ between
the atom and the cavity mode changes fast if the atoms pass the cavity within
a predefined time. So now we take into account N atom are trapped in a bimodal
cavity as showed in fig. 1a, and the trapping time has been demonstrated up to
several seconds \cite{rempe,kimble}. The two cavity modes have different
polarization $L$ (left-circularly polarized) and $R$ (right-circularly
polarized). The cavity photons leaked out through the mirror, can be detected
by two single-photon detectors $D_{0}$ and $D_{1}$ without failure, and
$D_{0}$ ($D_{1}$) is triggered by $L$ ($R$) photon due to the quarter-wave
plate (QWP). The trapped atoms have the identical five-level configuration, as
depicted in fig. 1b, the ground states $\left\vert 0\right\rangle $,
$\left\vert 1\right\rangle $, $\left\vert 2\right\rangle $ and the excited
state $\left\vert e\right\rangle $, $\left\vert r\right\rangle $. The
configuration can be obtained in $^{87}$Rb. For example, the states
$\left\vert 0\right\rangle $, $\left\vert 1\right\rangle $, and $\left\vert
2\right\rangle $, are respectively $\left\vert F=2,m=-1\right\rangle $,
$\left\vert F=2,m=1\right\rangle $, $\left\vert F=1,m=-1\right\rangle $ of
$5^{2}S_{1/2}$; $\left\vert e\right\rangle $ and $\left\vert r\right\rangle $
are respectively $\left\vert F=1,m=0\right\rangle $, $\left\vert
F=1,m=-1\right\rangle $ of $5^{2}P_{1/2}$. Atomic transition $\left\vert
0\right\rangle \longleftrightarrow\left\vert e\right\rangle $ ($\left\vert
1\right\rangle \longleftrightarrow\left\vert e\right\rangle $) is coupled with
cavity mode $a_{L}$ ($a_{R}$). The two $\pi$-polarized classical laser pulses
$\varepsilon_{1}$ and $\varepsilon_{2}$ are used to transfer the occupation of
the state $\left\vert 0\right\rangle $ to $\left\vert 2\right\rangle $\ with
Rabi frequencies $\Omega_{1}\left(  t\right)  $ and $\Omega_{2}\left(
t\right)  $ respectively.\begin{figure}[ptb]
\includegraphics[scale=1.0]{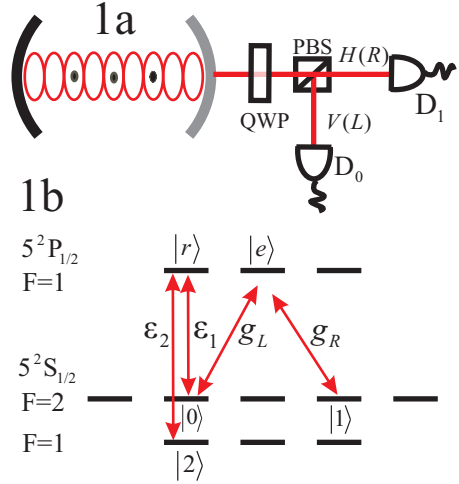}\caption{(a) Schematic setup to realize
multi-atom entangled state in a leaky optical cavity. QWP is a quarter-wave
plate, PBS is a polarization beamsplitter, and $D_{0}$ and $D_{1}$ are two
single-photon detectors. (b) The energy level diagram of atoms. The states
$\left\vert 0\right\rangle $, $\left\vert 1\right\rangle $\ and $\left\vert
2\right\rangle $\ are the hyperfine states in the ground-state manifold,
respectively; $\left\vert e\right\rangle $ and $\left\vert r\right\rangle $
are the excited state. Atomic transition $\left\vert 0\right\rangle
\longleftrightarrow\left\vert e\right\rangle $ ($\left\vert 1\right\rangle
\longleftrightarrow\left\vert e\right\rangle $) is coupled with the cavity
mode $a_{L}$ ($a_{R}$).}%
\end{figure}

The n atoms, expected to entangle together, can be arbitrarily chosen from N
trapped atoms. We first completely transfer the occupations of the state
$\left\vert 0\right\rangle $ of the other (N-n) atoms to those of the state
$\left\vert 2\right\rangle $. Then the (N-n) atoms are beyond systemic state
evolution space (cavity + n atoms). Before study the system evolution
dynamics, we do some consideration. The separation between any two atoms is
large enough compared with the wavelength of the fields in interest so that
the dipole-dipole interaction among atoms can be neglected. This assumption is
feasible because cavity length is always much larger than the wavelength of
cavity modes. We neglect spontaneous emission of the atoms. Subject to no
decay being recorded in the detectors, under the rotating wave approximation,
the system conditionally evolves according to a non-Hermitian Hamiltonian
which is given by (in units of $\hbar=1$)
\begin{align}
H_{1}  &  =\sum_{i=1}^{n}\left(  \omega_{e}\left\vert e\right\rangle
_{ii}\left\langle e\right\vert +\omega_{0}\left\vert 0\right\rangle
_{ii}\left\langle 0\right\vert +\omega_{1}\left\vert 1\right\rangle
_{ii}\left\langle 1\right\vert \right)  +\left(  \omega_{L}-\frac{i\kappa_{L}%
}{2}\right)  a_{L}^{\dagger}a_{L}\nonumber\\
&  +\left(  \omega_{R}-\frac{i\kappa_{R}}{2}\right)  a_{R}^{\dagger}a_{R}%
+\sum_{i=1}^{n}\left(  g_{L}\left\vert e\right\rangle _{ii}\left\langle
0\right\vert a_{L}+g_{R}\left\vert e\right\rangle _{ii}\left\langle
1\right\vert a_{R}+H.c.\right)  , \label{1th}%
\end{align}
where $\omega_{e}$ ($\omega_{0}$, $\omega_{1}$) is the atomic energy in the
atomic state $\left\vert e\right\rangle $ ($\left\vert 0\right\rangle $,
$\left\vert 1\right\rangle $); $\omega_{\mu=L,R}$ is the frequency for the
cavity mode $a_{\mu}$; $g_{\mu}$ (assumed real) is the single-photon coherent
coupling rate with the cavity mode $a_{\mu}$; $\kappa_{L}=\omega/Q_{L}$ and
$\kappa_{R}=\omega/Q_{R}$ denote the rate of decay of the cavity mode fields
$a_{L}$ and $a_{R}$, respectively, and $H.c.$ stands for the Hermitian
conjugate. Then the interaction Hamiltonian in the interaction picture can be
written as
\begin{equation}
H_{2}=\sum_{i=1}^{n}\left(  g_{L}\left\vert e\right\rangle _{ii}\left\langle
0\right\vert a_{L}+g_{R}\left\vert e\right\rangle _{ii}\left\langle
1\right\vert a_{R}+H.c.\right)  -\left(  \Delta_{L}+\frac{i\kappa_{L}}%
{2}\right)  a_{L}^{\dagger}a_{L}-\left(  \Delta_{R}+\frac{i\kappa_{R}}%
{2}\right)  a_{R}^{\dagger}a_{R}. \label{2th}%
\end{equation}
Here, the detuning $\Delta_{L}$ ($\Delta_{R}$) is defined as $\omega
_{e}-\omega_{0}-\omega_{L}$ ($\omega_{e}-\omega_{1}-\omega_{R}$), and our
scheme works in the strongly detuned limit $\Delta_{\mu}\gg g_{\mu^{\prime}}$.

Expressing the state of the total system in the form of $\left\vert
\text{atoms}\right\rangle \left\vert \text{cavity}\right\rangle $, the initial
state of the system is prepared as $\left\vert 0_{1}0_{2}\cdots0_{n}%
\right\rangle \left\vert L\right\rangle $, which means all atoms are in the
ground state $\left\vert 0\right\rangle $ and the cavity fields have a $L$
polarized photon. The temporal evolution of the system is spanned by the
$2n+1$ basis states: $\{\left\vert 0_{1}0_{2}\cdots0_{n}\right\rangle
\left\vert L\right\rangle $, $\left\vert e_{1}0_{2}\cdots0_{n}\right\rangle
\left\vert vac\right\rangle $, $\cdots$, $\left\vert 0_{1}0_{2}\cdots
e_{n}\right\rangle \left\vert vac\right\rangle $, $\left\vert 1_{1}0_{2}%
\cdots0_{n}\right\rangle \left\vert R\right\rangle $, $\cdots$, $\left\vert
0_{1}0_{2}\cdots1_{n}\right\rangle \left\vert R\right\rangle \}$. Since the
initial atomic state is symmetrical and the atoms are indistinguishable in the
evolution state space, the basis states reduce to $\{\left\vert \phi
_{0}\right\rangle =\left\vert 0_{1}0_{2}\cdots0_{n}\right\rangle \left\vert
L\right\rangle $, $\left\vert \phi_{1}\right\rangle =\frac{1}{\sqrt{n}}\left(
\left\vert 1_{1}0_{2}\cdots0_{n}\right\rangle +\cdots+\left\vert 0_{1}%
0_{2}\cdots1_{n}\right\rangle \right)  \left\vert R\right\rangle $,
$\left\vert \phi_{2}\right\rangle =\frac{1}{\sqrt{n}}\left(  \left\vert
e_{1}0_{2}\cdots0_{n}\right\rangle +\cdots+\left\vert 0_{1}0_{2}\cdots
e_{n}\right\rangle \right)  \left\vert vac\right\rangle \}$. The state of the
system at arbitrary time is described by
\begin{equation}
\left\vert \Psi\left(  t\right)  \right\rangle =c_{0}\left(  t\right)
\left\vert \phi_{0}\right\rangle +c_{1}\left(  t\right)  \left\vert \phi
_{1}\right\rangle +c_{2}\left(  t\right)  \left\vert \phi_{2}\right\rangle .
\label{3th}%
\end{equation}
According to Schr\"{o}dinger equation $i\partial_{t}\left\vert \Phi\left(
t\right)  \right\rangle =H_{2}\left\vert \Phi\left(  t\right)  \right\rangle
$, we have%
\begin{align}
idc_{0}\left(  t\right)  /dt  &  =-\left(  \Delta_{L}+\frac{i\kappa_{L}}%
{2}\right)  c_{0}\left(  t\right)  +\sqrt{n}g_{L}c_{2}\left(  t\right)
,\nonumber\\
idc_{1}\left(  t\right)  /dt  &  =-\left(  \Delta_{R}+\frac{i\kappa_{R}}%
{2}\right)  c_{1}\left(  t\right)  +g_{R}c_{2}\left(  t\right)  ,\label{4th}\\
idc_{2}\left(  t\right)  /dt  &  =\sqrt{n}g_{L}c_{0}\left(  t\right)
+g_{R}c_{1}\left(  t\right)  .\nonumber
\end{align}
Then we utilize the transformation $\lambda_{\nu=0,1,2}\left(  t\right)
=c_{\nu}\left(  t\right)  e^{-i\Delta_{L}t}$, therefore, eqs. $\left(
4\right)  $ can be written as%
\begin{align}
d\lambda_{0}\left(  t\right)  /dt  &  =-\frac{\kappa_{L}}{2}\lambda_{0}%
-i\sqrt{n}g_{L}\lambda_{2}\left(  t\right)  ,\nonumber\\
d\lambda_{1}\left(  t\right)  /dt  &  =-i\left[  \left(  \Delta_{L}-\Delta
_{R}\right)  -\frac{i}{2}\kappa_{R}\right]  \lambda_{1}\left(  t\right)
-ig_{R}\lambda_{2}\left(  t\right)  ,\label{5th}\\
d\lambda_{2}\left(  t\right)  /dt  &  =-i\sqrt{n}g_{L}\lambda_{0}\left(
t\right)  -ig_{R}\lambda_{1}\left(  t\right)  -i\Delta_{L}\lambda_{2}\left(
t\right)  .\nonumber
\end{align}
Assuming $g_{R}=g_{L}=g$, $\kappa_{L}=\kappa_{R}=\kappa$, $\Delta_{L}%
,\Delta_{R}\gg g$, and $\left\vert \Delta_{L}-\Delta_{R}\right\vert \ll g$,
the atomic population in the excited state $\left\vert e\right\rangle $ is
very small, and we can put $d\lambda_{2}\left(  t\right)  /dt=0$ \cite{biswas}
so that eqs. $\left(  3\right)  $ and $\left(  5\right)  $ reduce to%
\begin{align}
\left\vert \Psi\left(  t\right)  \right\rangle  &  =\lambda_{0}\left(
t\right)  \left\vert \phi_{0}\right\rangle +\lambda_{1}\left(  t\right)
\left\vert \phi_{1}\right\rangle ,\nonumber\\
d\lambda_{0}\left(  t\right)  /dt  &  =\left(  \frac{ing^{2}}{\Delta_{L}%
}-\frac{\kappa}{2}\right)  \lambda_{0}\left(  t\right)  +i\frac{\sqrt{n}g^{2}%
}{\Delta_{L}}\lambda_{1}\left(  t\right)  ,\label{6th}\\
d\lambda_{1}\left(  t\right)  /dt  &  =\frac{i\sqrt{n}g^{2}}{\Delta_{L}%
}\lambda_{0}\left(  t\right)  +\left[  \frac{ig^{2}}{\Delta_{L}}-i\left(
\Delta_{L}-\Delta_{R}\right)  -\frac{\kappa}{2}\right]  \lambda_{1}\left(
t\right)  .\nonumber
\end{align}
Here we neglect the common phase $e^{i\Delta_{L}t}$. The solutions of eqs.
$\left(  6\right)  $ are%
\begin{align}
\lambda_{0}\left(  t\right)   &  =e^{-\kappa t/2}e^{\frac{i\Omega_{0}t}{2}%
}\left[  \cos\left(  \frac{\Omega_{1}t}{2}\right)  +i\frac{2ng^{2}-\Delta
_{L}\Omega_{0}}{\Delta_{L}\Omega_{1}}\sin\left(  \frac{\Omega_{1}t}{2}\right)
\right]  ,\nonumber\\
\lambda_{1}\left(  t\right)   &  =ie^{-\kappa t/2}e^{\frac{i\Omega_{0}t}{2}%
}\frac{\Delta_{L}^{2}\Omega_{1}^{2}-\left(  \Delta_{L}\Omega_{0}%
-2ng^{2}\right)  ^{2}}{2\sqrt{n}g^{2}\Delta_{L}\Omega_{1}}\sin\left(
\frac{\Omega_{1}t}{2}\right)  , \label{7th}%
\end{align}
where $\Omega_{0}=\frac{\left(  n+1\right)  g^{2}}{\Delta_{L}}-\left(
\Delta_{L}-\Delta_{R}\right)  $, $\Omega_{1}=\sqrt{\left[  \frac{\left(
n+1\right)  g^{2}}{\Delta_{L}}\right]  ^{2}+\left(  \Delta_{L}-\Delta
_{R}\right)  ^{2}+2\left(  n-1\right)  \left(  \Delta_{L}-\Delta_{R}\right)
\frac{g^{2}}{\Delta_{L}}}$. When $n=1$ and $\kappa=0$, it is the result of ref
\cite{biswas}. In order to obtain multi-atom entangled state,\ we consider the
case of $n>1$. If we have%
\begin{equation}
\Delta_{L}-\Delta_{R}=\left(  1-n\right)  g^{2}/\Delta_{L} \label{8th}%
\end{equation}
and thus
\begin{align}
\Omega_{0}  &  =2ng^{2}/\Delta_{L},\nonumber\\
\Omega_{1}  &  =2\sqrt{n}g^{2}/\Delta_{L}, \label{9th}%
\end{align}
Then eqs. $\left(  7\right)  $ reduce to a laconic expression%
\begin{align}
\lambda_{0}\left(  t\right)   &  =e^{-\kappa t/2}e^{\frac{i\Omega_{0}t}{2}%
}\cos\left(  \frac{\Omega_{1}t}{2}\right)  ,\nonumber\\
\lambda_{1}\left(  t\right)   &  =e^{-\kappa t/2}ie^{\frac{i\Omega_{0}t}{2}%
}\sin\left(  \frac{\Omega_{1}t}{2}\right)  . \label{10th}%
\end{align}
The factor $e^{-\kappa t/2}$ describes the leakage of cavity photons.

Assume the detector $D_{0}$ or $D_{1}$ has a response at time $t=t_{r}\leq T$
(Here, $T$\ is a waiting time of two detectors), and then this coherent time
evolution $\left\vert \Psi\left(  t\right)  \right\rangle $ governed by
$H_{2}$ is immediately interrupted by a corresponding quantum jump operator
\cite{jump} $b_{0}\left(  =a_{L}\right)  $ or $b_{1}\left(  =a_{R}\right)  $,
respectively. After the detector $D_{k}$ responses, the state of system can be
written as%
\begin{equation}
\left\vert \Psi\left(  t_{r}\right)  \right\rangle ^{D_{k}}=\frac
{b_{k}\left\vert \Psi\left(  t_{r}\right)  \right\rangle }{\left\Vert
b_{k}\left\vert \Psi\left(  t_{r}\right)  \right\rangle \right\Vert }.
\label{11th}%
\end{equation}

In the case that $D_{1}$ responses, after tracing out the cavity modes part,
we can see that the n atoms are prepared in the entangled state $\left\vert
\varphi\right\rangle _{ent}=\frac{1}{\sqrt{n}}\left(  \left\vert 1_{1}%
0_{2}\cdots0_{n}\right\rangle +\cdots+\left\vert 0_{1}0_{2}\cdots
1_{n}\right\rangle \right)  $. $\left\vert \varphi\right\rangle _{ent}$ is
right a W state \cite{dur} of n particles, which is a special case ($m=1$) of
Dicke states $\left\vert n,m\right\rangle _{Dicke}$. Notice that if we let
$\sqrt{n}g_{L}=g_{R}$ in eqs. $\left(  4\right)  $ and $\left(  5\right)  $,
the original multi-atom model reduces to a symmetric three-level system in the
ref. \cite{biswas}. In this case, two-photon resonant condition $\Delta
_{L}=\Delta_{R}$ should be met in order to generate $\left\vert \varphi
\right\rangle _{ent}$. But the condition of $\sqrt{n}g_{L}=g_{R}$ is not
always satisfied\ in actual system because $g_{L}/g_{R}$ has a determinate
relation for two given atomic transitions. Obviously, the state of the n atoms
relapses to their initial state $\left\vert 0_{1}0_{2}\cdots0_{n}\right\rangle
$ if $D_{0}$ is triggered. With the reinjection of a new left-circularly
polarized photon into the cavity, the same process is repeated until the
entangled state is prepared. Similar to the ref. \cite{hong}, we can make the
experiment automatically repeat itself through\ replacing the detector $D_{0}$
by a path directed back to the cavity, so that the left-circularly polarized
photon can be automatically fed back to the cavity. Therefore, $\left\vert
\varphi\right\rangle _{ent}$ can be produced with near unit probability.
Furthermore, if we reinject a new left-circularly polarized photon into the
cavity after a W state of n atoms is achieved, we can prepare Dicke state
$\left\vert n,2\right\rangle _{Dicke}$. Moreover, if the same process is
repeated, universal n-qubit Dicke states $\left\vert n,m\right\rangle
_{Dicke}$ can be prepared step by step \cite{yu}.

Now we turn to analyze the efficiency of the scheme. In the absence of
spontaneous emission of the atoms, From eqs. $\left(  10\right)  $, one can
know that the probability of photons decay from cavity is $P_{decay}\left(
t\right)  =1-\left\vert e^{-\kappa t/2}\right\vert ^{2}$, and in the time
interval $dt_{r}$, the\ probability that the detector $D_{1}$ is triggered is
$\left\vert \sin\left(  \frac{\Omega_{1}t}{2}\right)  \right\vert ^{2}%
\cdot\left(  \frac{dP_{decay}\left(  t_{r}\right)  }{dt_{r}}\right)  dt_{r}$.
Noticeably, a left- or right-circularly polarized photon will finally leak out
due to the factor $e^{-\kappa t/2}$ in eqs. $\left(  10\right)  $. In fact, in
the case of $\kappa T\gg1$, we consider one of detectors should response with
the time $T$, so that the\ probability of the detector $D_{1}$ is triggered,
i.e., the success probability $p_{suc}$ of entangled state $\left\vert
\varphi\right\rangle _{ent}$ being prepared, can be roughly calculated as%
\begin{align}
p_{suc}  &  \approx%
{\displaystyle\int\nolimits_{0}^{T\gg1/\kappa}}
\left\vert \sin\left(  \frac{\Omega_{1}t}{2}\right)  \right\vert ^{2}%
\cdot\left(  \frac{dP_{decay}\left(  t_{r}\right)  }{dt_{r}}\right)
dt_{r}\nonumber\\
&  \approx\frac{2ng^{4}}{\Delta_{L}^{2}\left(  \kappa^{2}+\Omega_{1}%
^{2}\right)  }. \label{12th}%
\end{align}
In order to get the maximum of $p_{suc}$, we find that it is corresponds to
the minimum of $\Omega_{1}$, that is, $\left(  \Delta_{L}-\Delta_{R}\right)
=-\frac{\left(  n-1\right)  g^{2}}{\Delta_{L}}$. Especially take notice that
this condition agrees with eqs. $\left(  8,9\right)  $. So we have%
\begin{equation}
p_{suc}=\frac{2ng^{4}}{\Delta_{L}^{2}\kappa^{2}+4ng^{4}}. \label{13th}%
\end{equation}
As showed in fig. 2, we can find, for a given $\Delta_{L}$, iff $\kappa\ll g$,
$p_{suc}\left(  g,\kappa\right)  $ reaches near $50\%$.\begin{figure}[ptb]
\includegraphics[scale=1.0]{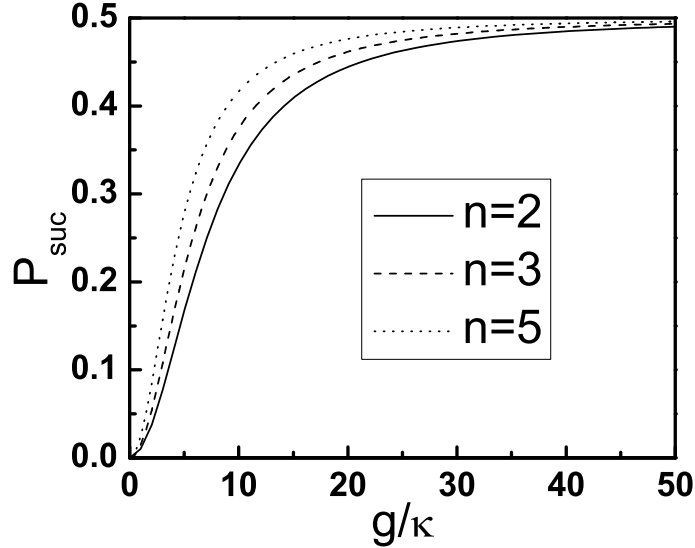}\caption{The success probability
$p_{suc}$ computed numerically as a function $g/\kappa$ and $n$. Other
parameter: $\Delta_{L}=20g$.}%
\end{figure}

We consider a set of practical parameters. $g=2\pi\times16$ $%
\operatorname{MHz}%
$ and $\kappa=2\pi\times1.4$ $%
\operatorname{MHz}%
$ are given in Rempe's group \cite{rempe}. We also choose $T=0.5$ $%
\operatorname{\mu s}%
$ (Commonly, the coherence time of atomic internal states in a high-$Q$ cavity
is much larger than the order of $T$ \cite{kuhr}), $\Delta_{L}=20g$, and
$n=3$. After a trial, $p_{suc}=\frac{2ng^{4}}{\Delta_{L}^{2}\kappa^{2}%
+4ng^{4}}\sim0.36$. The probability is not very high, but it increases to
quai-unit after several trials. For instance, the total success probability is
more than $99\%$ after ten trials.

In addition, the atoms always\ has spontaneous emission when they are in the
excited states. However, the occupation of the excited states is very small
since the atomic transitions are largely detuned with cavity modes in our
scheme. Our numerical simulation shows the occupation is less than $3\%$ based
on above practical parameters. Therefore, atomic spontaneous emission is an
inappreciable ingredient in our scheme.

F-P cavity has a general\ mode function described by $\chi\left(
\overrightarrow{r}\right)  =\sin\left(  kz\right)  \exp[-\left(  x^{2}%
+y^{2}\right)  /w_{0}^{2}]$ \cite{duan2}, and $g\left(  r\right)  =g_{0}%
\chi\left(  \overrightarrow{r}\right)  $, where $w_{0}$ and $k=2\pi/\lambda$
are, respectively, the width and the wave vector of the Gaussian cavity mode,
and $\overrightarrow{r}\left(  x,y,z\right)  $ describes the atomic locations;
$z$ is assumed to be along the axis of the cavity. In the above description we
have assumed the coupling rate $g_{L,R}^{i}\left(  r\right)  =g$, which means
all atoms are addressed on some certain locations. An obviously best case is:
$\sin\left(  k_{L,R}z\right)  =1$ and $x=y=0$. Apparently, these conditions
still remain a challenge based on current cavity QED technology. We also note
our scheme requires an efficient source of single photons \cite{sps} and their
injection into\ an optical cavity. However, as a theoretical design, our
scheme is still\ potentially feasible in the near future technology, for all
these obstacles are now under active investigations \cite{kimble,mckeever} and
may be overcame in the near future.

In conclusion, we have presented a scheme to prepare multi-atom entangled
state in an optical cavity. Based on current and near future cavity QED
technology, considering and using the dissipation of the system, we design a
scheme based on measurements to leaky photons, to generate universal Dicke
states $\left\vert n,m\right\rangle _{Dicke}$ of n atoms. In principle,
although the preparation is probabilistic in one trial, the probability of
success reaches\ close to unit after several trials. Obviously, our scheme has
ultra-high fidelity, since non-ideal single-photon detectors, absorption by
cavity mirrors and atomic spontaneous emission just decrease the success
probability, not the fidelity of states.

\begin{acknowledgments}
We thank Professor Y.-S. Zhang and X.-M. Lin for their helpful advice. This
work was funded by National Fundamental Research Program of China (Grant No.
2001CB309300) and the Innovation Funds of the Chinese Academy of Sciences.
\end{acknowledgments}

\end{document}